\documentclass[fleqn,usenatbib]{mnras}

\usepackage{newtxtext,newtxmath}

\usepackage[T1]{fontenc}

\DeclareRobustCommand{\VAN}[3]{#2}
\let\VANthebibliography\thebibliography
\def\thebibliography{\DeclareRobustCommand{\VAN}[3]{##3}\VANthebibliography}

\usepackage{graphicx}	%
\usepackage{amsmath}	%

\newcommand{\be}{\begin{equation}}
\newcommand{\ee}{\end{equation}}
\newcommand{\beqa}{\begin{eqnarray}}
\newcommand{\eeqa}{\end{eqnarray}}

\title[LISA constraints on IMBH in GC]{LISA Constraints on an Intermediate-Mass Black Hole\\ in the Galactic Centre}

\author[Strokov, Fragione, Berti]{
Vladimir Strokov,$^{1}$\thanks{E-mail: vstroko1@jhu.edu}
Giacomo Fragione,$^{2,3}$
and Emanuele Berti$^{1}$
\\
$^{1}$Johns Hopkins University, 3400 North Charles Street, Baltimore, MD 21218, USA\\
$^{2}$Center for Interdisciplinary Exploration \& Research in Astrophysics (CIERA), Northwestern University, 1800 Sherman Ave, Evanston, IL 60201, USA\\
$^{3}$Department of Physics \& Astronomy, Northwestern University, Evanston, IL 60208, USA
}

\date{Accepted XXX. Received YYY; in original form ZZZ}

\pubyear{2023}

\begin{document}
\label{firstpage}
\pagerange{\pageref{firstpage}--\pageref{lastpage}}
\maketitle

\begin{abstract}
Galactic nuclei are potential hosts for intermediate-mass black holes (IMBHs), whose gravitational field can affect the motion of stars and compact objects. The absence of observable perturbations in our own Galactic Centre has resulted in a few constraints on the mass and orbit of a putative IMBH. Here, we show that the Laser Interferometer Space Antenna (LISA) can further constrain these parameters if the IMBH forms a binary with a compact remnant (a white dwarf, a neutron star, or a stellar-mass black hole), as the gravitational-wave signal from the binary will exhibit Doppler-shift variations as it orbits around Sgr~A$^*$. We argue that this method is the most effective for IMBHs with masses $10^3\,M_\odot\lesssim M_{\rm IMBH}\lesssim 10^5\,M_\odot$ and distances of $0.1$--$2$~mpc with respect to the supermassive black hole, a region of the parameter space partially unconstrained by other methods. We show that in this region the Doppler shift is most likely measurable whenever the binary is detected in the LISA band, and it can help constrain the mass and orbit of a putative IMBH in the centre of our Galaxy. We also discuss possible ways for an IMBH to form a binary in the Galactic Centre, showing that gravitational-wave captures of stellar-mass black holes and neutron stars are the most efficient channel.
\end{abstract}


\section{Introduction} \label{sec:intro}

Intermediate-mass black holes (IMBHs) are among the most elusive astrophysical objects. As suggested by their name, IMBHs occupy a mass range between stellar black holes, with masses $\lesssim 100M_\odot$\,, and supermassive black holes (SMBHs), with masses~$\gtrsim 10^6M_\odot$\,. Although without inconclusive final evidence, IMBHs have been hunted for in a variety of ways~\citep{Greene:2019vlv}. Much like their less and more massive counterparts, IMBHs can undergo a phase of accretion and become visible in the electromagnetic spectrum~\citep[e.g.,][]{Greene:2006kg,Greene:2007wy,Kaaret:2017tcn,Chilingarian:2018acs,2018ApJ...868..152B}, tidally disrupt nearby stars resulting in bright and long transients~\citep[e.g.,][]{Shen:2013oma, Lin:2018dev, Chen:2018foj,Fragione:2018lvy,Lin:2021tfy}, affect the velocity dispersion profiles of their host star clusters~\citep[e.g.,][]{vanderMarel:2009mc,Noyola:2010ab,2011A&A...533A..36L,2019MNRAS.488.5340B}, and form binaries that merge via gravitational-wave (GW) emission~\citep[e.g.,][]{Gair:2010dx,Jani:2019ffg,Fragione:2022ams}. IMBHs may also be revealed with pulsar timing~\citep[e.g.,][]{2012ApJ...752...67K,Prager:2016puh,2017Natur.542..203K} and microlensing~\citep[e.g.,][]{2016MNRAS.460.2025K}.

\begin{figure*}
  \centering
  \includegraphics[width=0.45\textwidth]{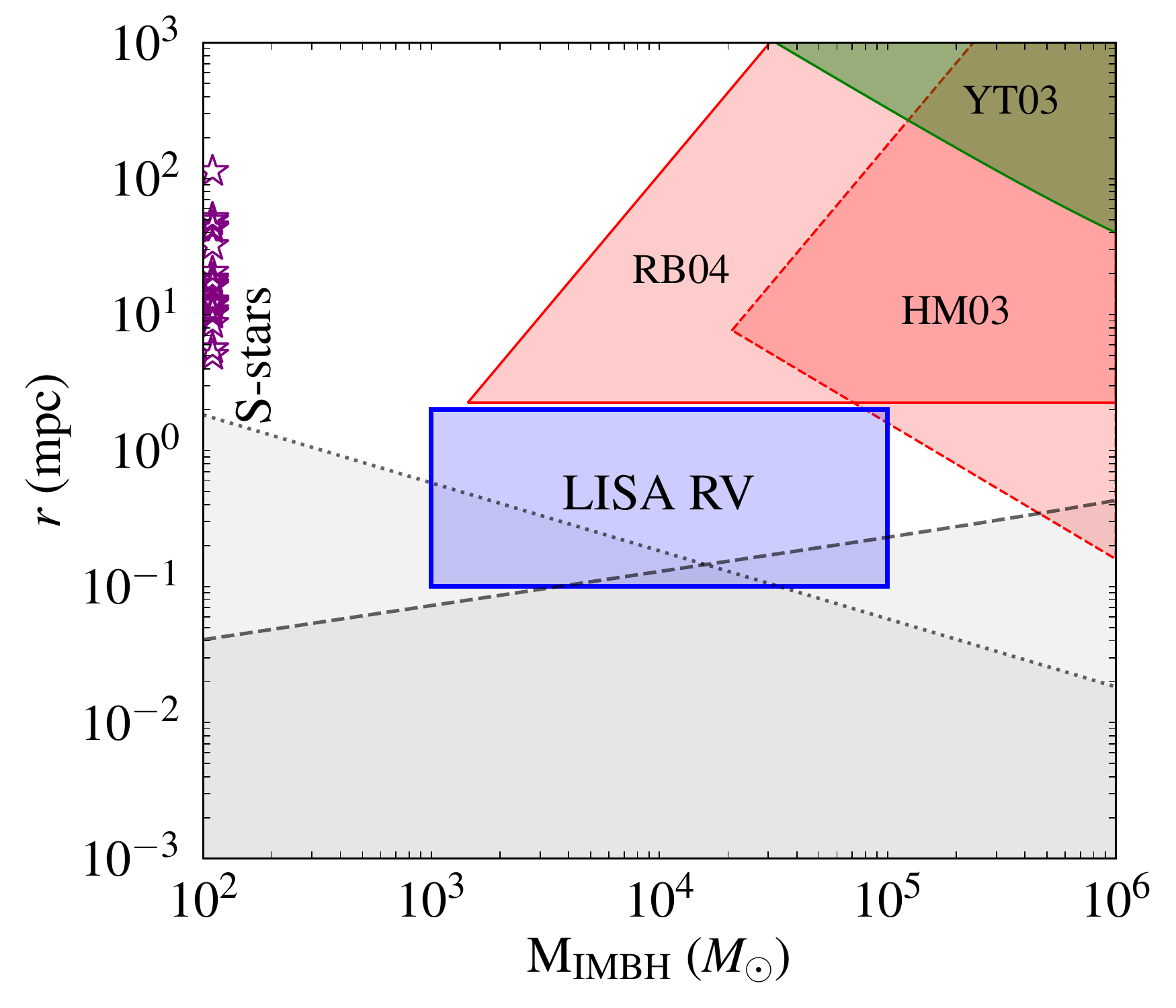}
  \includegraphics[width=0.45\textwidth]{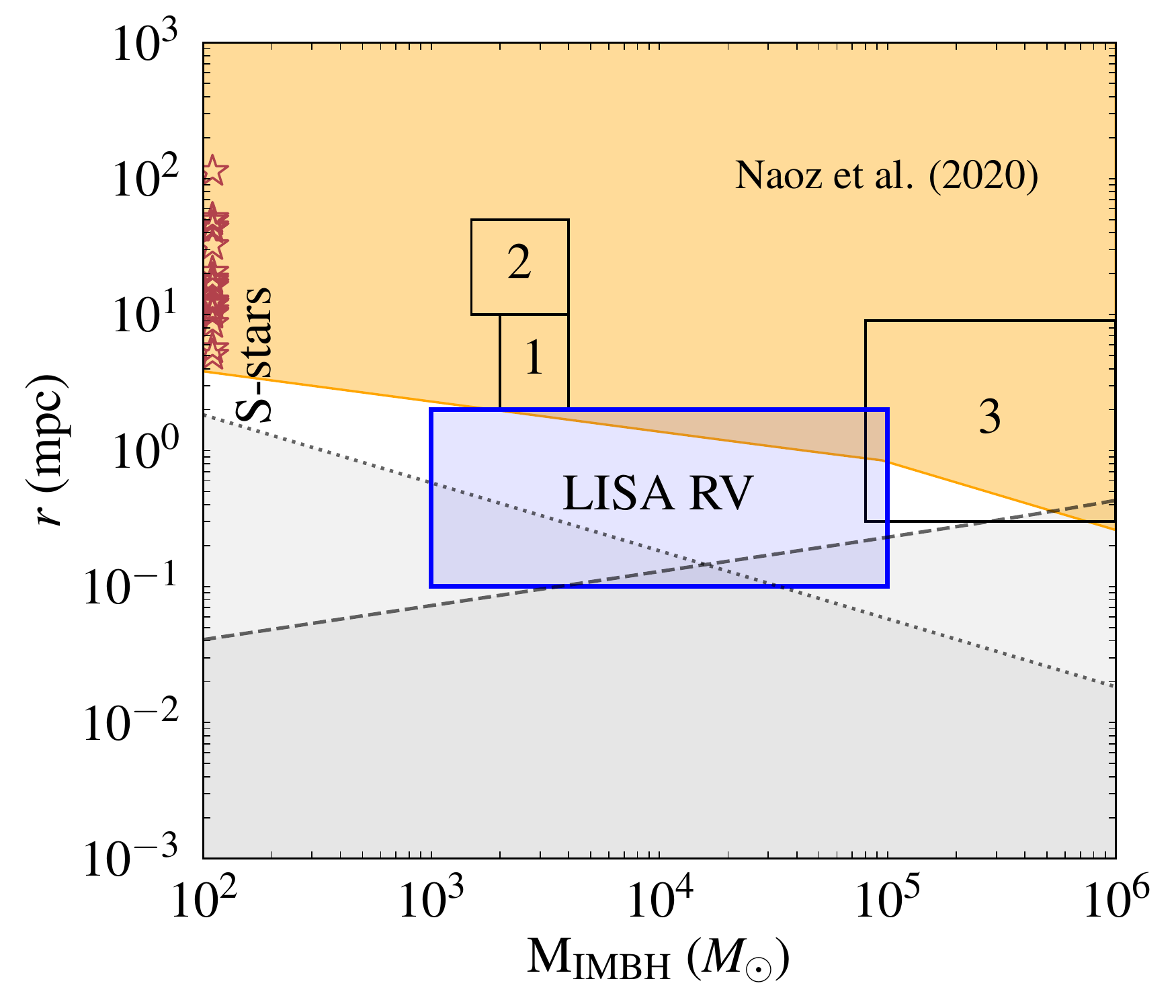}
  \caption{Constraints on the mass~$M_{\rm IMBH}$ of a potential IMBH in the Galactic Centre and its distance~$r$ from the central~SMBH. Left panel: Constraints from the velocity and position of Sgr~A$^\star$, where HM03, RB04, YT03 stand for \citet{Hansen:2003yb,Reid:2004rd,Yu:2003hj}, respectively. Right panel: Constraints from the motion of S-stars. The numbered boxes show portions of the parameter space excluded due to: (1) the effect of an IMBH on the distribution of separations and eccentricities of the S-stars~\citep{Gualandris:2009cc}, (2) efficient randomisation of the orbital inclinations~\citep{Merritt:2008ns}, and (3) extra mass within the orbit of S2~\citep{Gillessen:2008qv}. The orange shaded region represents the combined orbital stability and observational constraints by~\citet{Naoz:2019sjx}. Both panels: The purple stars on the vertical axis mark the semimajor axes of S-stars (plotted at $100\,M_\odot$ for illustrative purpose, but with typical mass of $\sim 10\,M_\odot$). The grey shaded regions represent limits on the inspiral time of the IMBH into Sgr~A$^\star$ ($t<1\;\mbox{Myr}$; below the dashed line) and that of the intermediate-mass ratio inspiral (IMRI) ($t<1\;\mbox{kyr}$; dotted line) under the assumption of circular orbits. In the IMRI case, the initial separation is assumed to be at the hard-soft boundary, Eq.~(\ref{eq:aIMRI}). The blue region labelled ``LISA RV'' shows the region of the parameter space we focus on in this paper.\label{fig:constraints}}%
\end{figure*}  

One of the potential sites where IMBHs can lurk are galactic centres. IMBHs can either be brought in there by disrupted globular clusters~\citep[e.g.,][]{Gnedin:2013cda,Fragione:2022egh} or form in situ~\citep[e.g.,][]{Rose:2021ftz}. For example, if an IMBH from a disrupted globular cluster drags in a handful of stars that remain bound to it, this could explain the population of young massive stars (S-stars) orbiting very close to Sgr~A$^\star$~\citep{Hansen:2003yb}. Our own Galactic Centre is the obvious first choice to search for an IMBH. A few constraints have been put on the mass $M_{\rm IMBH}$ of a potential IMBH and its distance~$r$ from the central SMBH. Those are based on non-detectability of various effects that the IMBH should dictate: a difference in the positions of the peaks of mass and light distribution in the Galactic Centre~\citep{Yu:2003hj}, peculiar velocity of the Sgr~A$^\star$ radio source~\citep{Hansen:2003yb,Reid:2004rd}, and perturbations to the orbits of the S-stars~\citep{Gillessen:2008qv,Merritt:2008ns,Gualandris:2009cc,Naoz:2019sjx,2023arXiv230108271Z}. Figure~\ref{fig:constraints} summarises the constraints; see also~\citet[Figure~13]{Gualandris:2009cc}, \citet[Figure~18]{Gillessen:2008qv}, and \citet[Figure~3]{Naoz:2019sjx}.

Owing to the high density of our Galactic Centre \citep{2018A&A...609A..26G,Schodel:2017vjf}, an IMBH may capture stars and compact objects while orbiting Sgr~A$^\star$. The resulting binary can merge via emission of GWs producing an IMRI. In this paper, we explore the possibility of detecting an IMBH in our Galactic Centre by using Doppler shift measurements in LISA~\citep{2017arXiv170200786A,LISA:2022yao}, and focus on a part of the $M_{\rm IMBH}$--$r$ parameter space so far unconstrained by other measurements (\mbox{$10^3M_\odot\lesssim M_{\rm IMBH}\lesssim 10^5M_\odot$} and \mbox{$0.1\;\mbox{mpc}<r<2\;\mbox{mpc}$}; see also Fig.~\ref{fig:constraints}).  We consider different cases for the mass of the secondary companion and estimate what fraction of the binaries could produce a GW Doppler shift detectable by LISA. We repeat our calculations under different assumptions on the distribution of separations between the IMBH and its companion. Finally, since we are looking to constrain the presence of an IMBH at $0.1$--$2\;\mbox{mpc}$ from Sgr~A$^\star$, we discuss possible scenarios for the formation of an IMRI at those distances.  

The paper is organized as follows. In Section~\ref{sec:fisher}, we elaborate on our assumptions and estimate the fraction of IMRIs with a detectable Doppler shift. Then, we present the various channels of IMRI formation in Section~\ref{sec:scenarios}. Finally, in Section~\ref{sec:discussion}, we discuss the implications of our results and how LISA can potentially contribute to constrain the presence of an IMBH in the Galactic Centre.

\section{Detection probability of Doppler shift}\label{sec:fisher}

If an IMBH forms an IMRI with a stellar-mass object while orbiting the central SMBH, LISA is well suited to detect the Doppler shift in the IMRI's GW signal. The IMRI is stable (``hard'') if its separation satisfies the condition~\citep{Sesana:2006xw}
\beqa
a_{\rm IMRI} \lesssim \frac{GM_{\rm IMBH}}{4v^2} &\simeq& 0.01\;\mbox{AU}\,\left(\frac{r}{1\;\mbox{mpc}}\right)\nonumber \\&\times&\left(\frac{M_{\rm IMBH}}{10^3M_\odot}\right)\left(\frac{M_{\rm SMBH}}{4\times 10^6M_\odot}\right)^{-1},
\label{eq:aIMRI}
\eeqa
where $v\sim\sqrt{GM_{\rm SMBH}/r}$ is the local velocity dispersion ($G$ is the gravitational constant). This semi-major axis corresponds to a GW frequency approximately in the middle of the LISA frequency band~\citep{2017arXiv170200786A,LISA:2022yao}
\be
f_{\rm IMRI} \simeq 2\;\mbox{mHz}\,\left(\frac{M_{\rm IMBH}}{10^3M_\odot}\right)^{1/2}\left(\frac{a}{0.01\;\mbox{AU}}\right)^{-3/2}\,.
\label{eq:fIMRI}
\ee
Moreover, for the range~\mbox{$0.1\;\mbox{mpc}<r<2\;\mbox{mpc}$}, the timescale for the variations of radial velocity
\be
T_{\rm RV}\simeq 2\pi\sqrt{\frac{r^3}{GM_{\rm SMBH}}} \approx 4\;\mbox{yr}\,\left(\frac{r}{2\;\mbox{mpc}}\right)^{3/2}\left(\frac{M_{\rm SMBH}}{4\times 10^6M_\odot}\right)^{-1/2}
\label{eq:TRV}
\ee
is shorter than the LISA observation time, which we assume to be $T_{\rm obs}=4$~yr~\citep{2017arXiv170200786A,2021arXiv210709665S,LISA:2022yao}. This ensures that at least one full Doppler-shift cycle can be traced if the RV amplitude is large enough~\citep{Randall:2018lnh}. Other applications of the Doppler shift method for GW signals include detecting a third body that perturbs the motion of a stellar-mass binary~\citep{Meiron:2016ipr,Inayoshi:2017hgw}.

The IMRIs we deal with are made up of a primary with mass in the IMBH range, $M_{\rm IMBH}\in[10^2\,M_\odot,10^5\,M_\odot]$, and a stellar-mass secondary. We consider three cases for the mass of the secondary: $m=0.6\,M_\odot$, $1.4\,M_\odot$, and $10\,M_\odot$\,, which approximately represent the typical masses of a white dwarf (WD), neutron star (NS), and a stellar-mass black hole (BH), respectively. We exclude main-sequence stars (MSs) as binary companions, because the typical semimajor axis given by Eq.~(\ref{eq:aIMRI}) is well within their tidal disruption radius. We assume that the distribution of the binaries' semimajor axes~$a$ follows a power law between $a_{\rm min}=0.01$~AU and~$a_{\rm max}=100$~AU, and we consider three values for the logarithmic slope: $\alpha=0.5$, $1$ (log-uniform), and~$1.5$. The orbital orientations of the IMRIs are sampled isotropically on a sphere, whereas the sky positions are fixed to the location of the Galactic Centre. For the distance of the IMRI from the central SMBH, we use a grid of values, namely $r\in[0.1,0.2,0.3,0.4,0.7,1,1.2,2]$~mpc. Following~\citet{Wong:2019hsq}, as detectability criteria, we require (i)~signal-to-noise ratio (SNR) higher than~$10$, and (ii)~relative uncertainty of both the radial velocity (RV) and period better than~$10\%$. The orbits of both the IMRI and the ``outer'' IMRI--Sgr~A$^\star$ binary are assumed to be circular (see Section~\ref{sec:discussion} for a discussion of the effects of eccentricity).

To estimate the uncertainties, we rely on the method of Fisher matrices~\citep{Vallisneri:2007ev} with a waveform for non-spinning BHs that includes corrections up to second post-Newtonian~(2PN) order~\citep[e.g.][]{Berti:2004bd}. Our calculation is based on the LISA sensitivity curve in~\citet{2021arXiv210801167B} adopted in the LISA Science Requirements Document (SciRD). For details of the Fisher matrix calculation and for the treatment of Doppler-shift corrections to the waveform, we refer the reader to~\citet[Section~II]{Strokov:2021mkv}.

\begin{figure}
  \centering
  \includegraphics[width=1.\columnwidth]{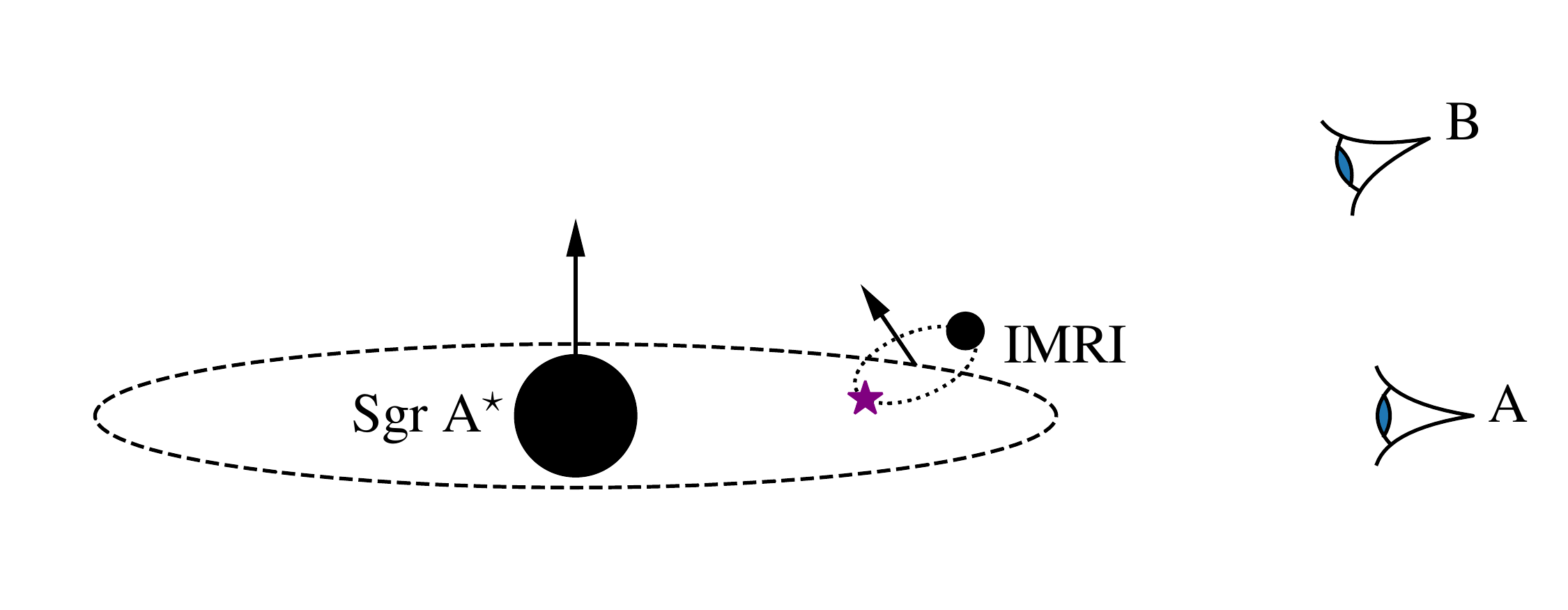}
  \caption{Illustration of the orbital configuration used in this paper. An IMBH (smaller black circle) forms an IMRI (``inner'' orbit) with a stellar-mass companion (purple star). The center of mass of the IMRI, in turn, follows an ``outer'' orbit around Sgr~A$^\star$ (bigger black circle). The arrows indicate the orientation of the two orbital angular momenta. The calculation of RV uncertainties is carried out assuming an edge-on ``outer'' orbit (as seen by observer~A on the right of the figure) and $56$~different orientations of the ``inner'' orbit. For observer~B (top right), who sees the ``outer'' orbit at a different inclination, the relative RV uncertainties are divided by the sine of the inclination to account for the RV amplitude--inclination degeneracy.\label{fig:picture}}%
\end{figure}

To speed up the calculation of the fraction of IMRIs with detectable Doppler shifts, we use the procedure illustrated in Fig.~\ref{fig:picture}.

First, we compute SNRs and the RV uncertainties on a $100\times 100$ logarithmic grid with $M_{\rm IMBH}\in[10^2M_\odot,10^5M_\odot]$ and~$a(\mbox{AU})\in[0.01, 1]$. We do so assuming an edge-on orbit of the IMRI around Sgr~A$^\star$ and $56$~different orientations of the IMRI's orbit (these form a uniform regular grid on a unit sphere). This computation is repeated for each choice of the mass of the secondary and for the values of~$r$ provided above. Second, we find the contours of $\mbox{SNR}=10$ and the RV uncertainty of~$10\%$ in the $M_{\rm IMBH}$--$a$ plane. We can neglect the uncertainty in the period, because in all cases that uncertainty is a few orders of magnitude below the one for the radial velocity. For a given IMBH mass, each contour now results in cutoff values~$a_{\rm SNR}$ and~$a_{\rm RV}$ for the semimajor axis. Then, given a power-law distribution of the semimajor axes, the fraction of detectable IMRIs can be computed analytically as an integral of the distribution with limits $a_{\rm min}$ and~$\min(a_{\rm SNR},a_{\rm RV})$. In case of an inclination of the IMRI--SMBH orbit different from~$90^\circ$ (see Fig.~\ref{fig:picture}), the contours for the edge-on RV uncertainty are trivially modified to account for the usual RV amplitude--inclination degeneracy. The cutoff values of~$a$ are modified accordingly. To quantify the effect of the inclination on the computed fraction, we use $100$~random isotropic inclinations for each value of~$M_{\rm IMBH}$.

\begin{figure}
  \centering
  \includegraphics[width=0.9\columnwidth]{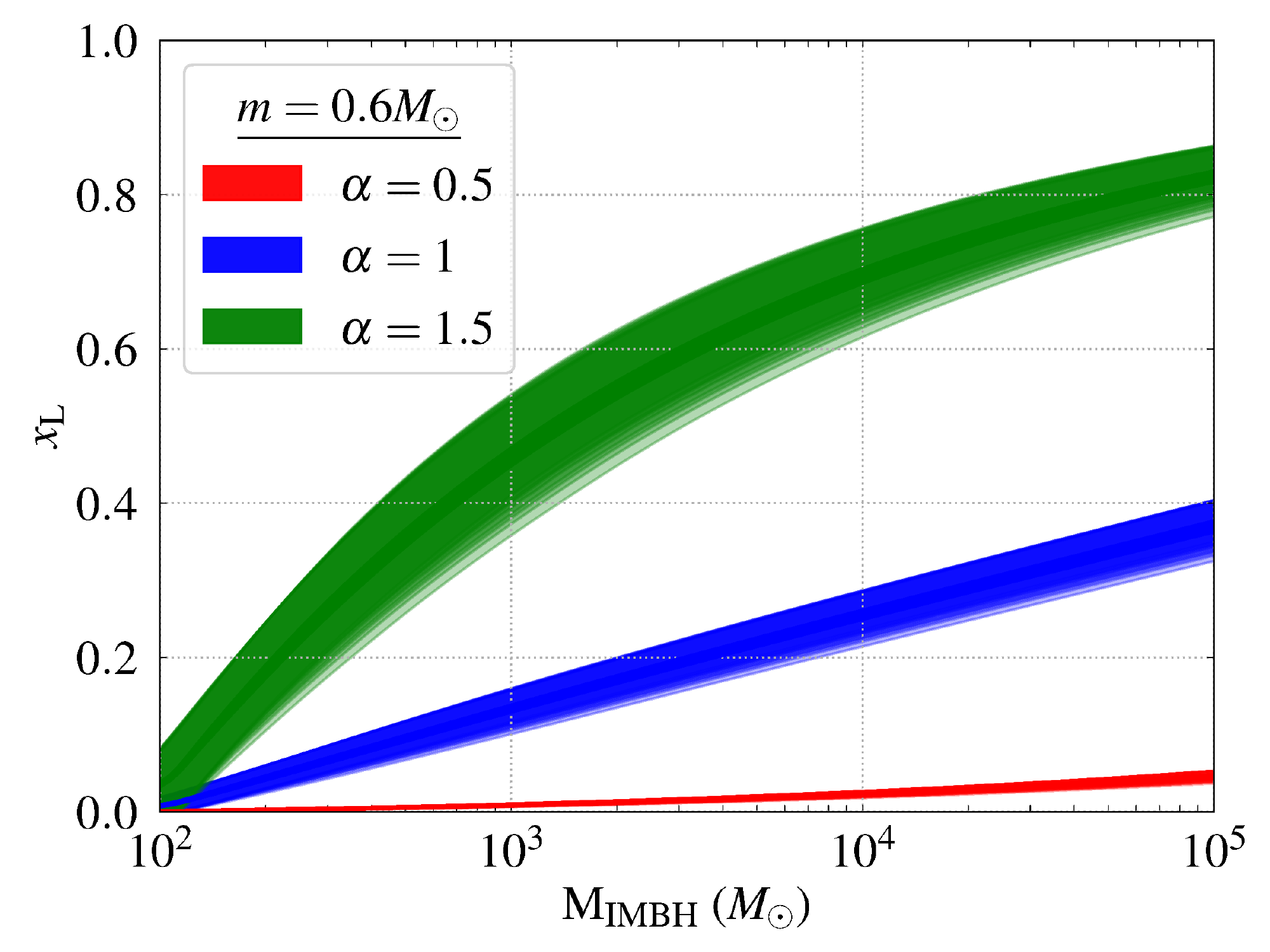}
  \includegraphics[width=0.9\columnwidth]{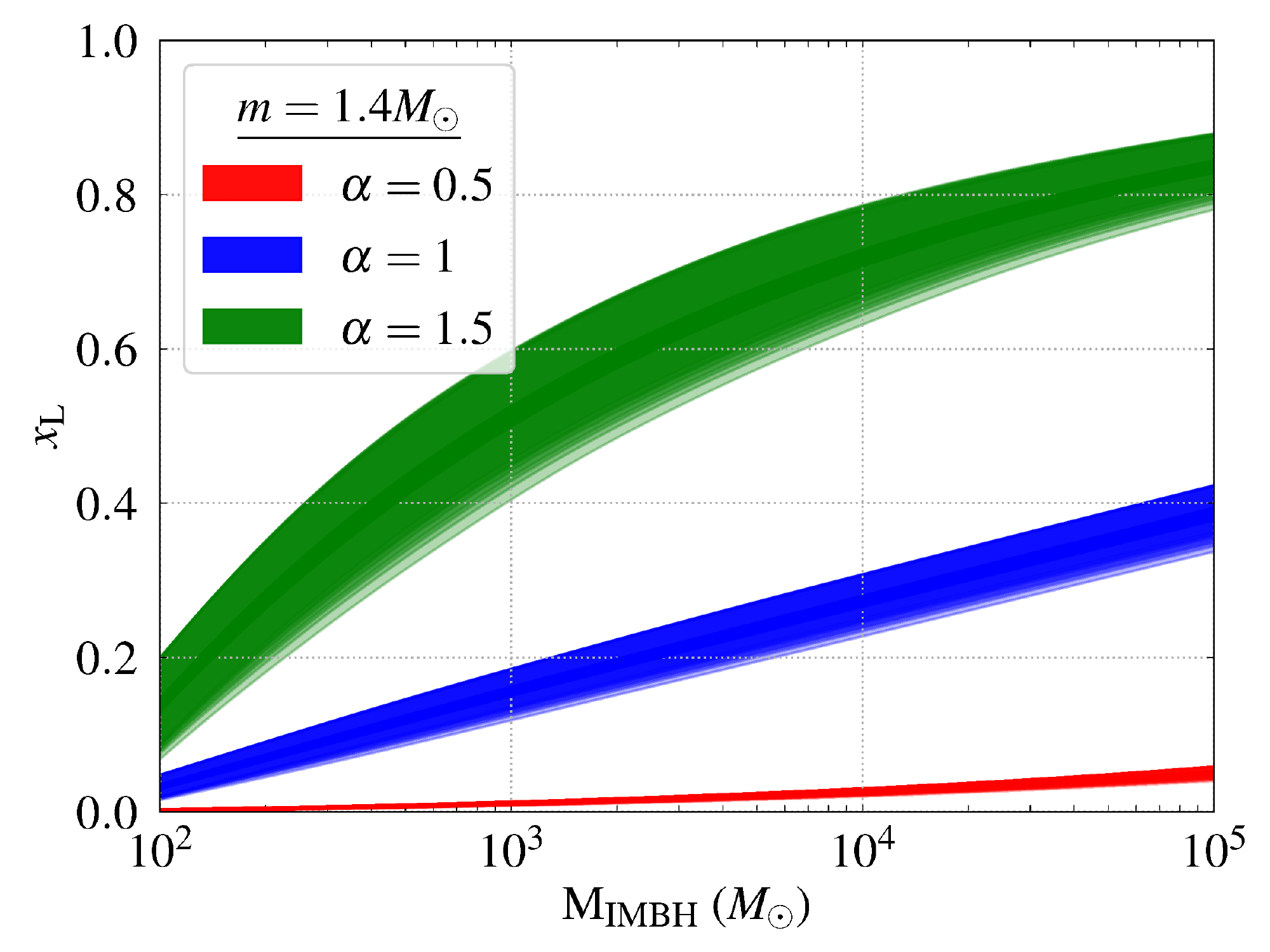}
  \includegraphics[width=0.9\columnwidth]{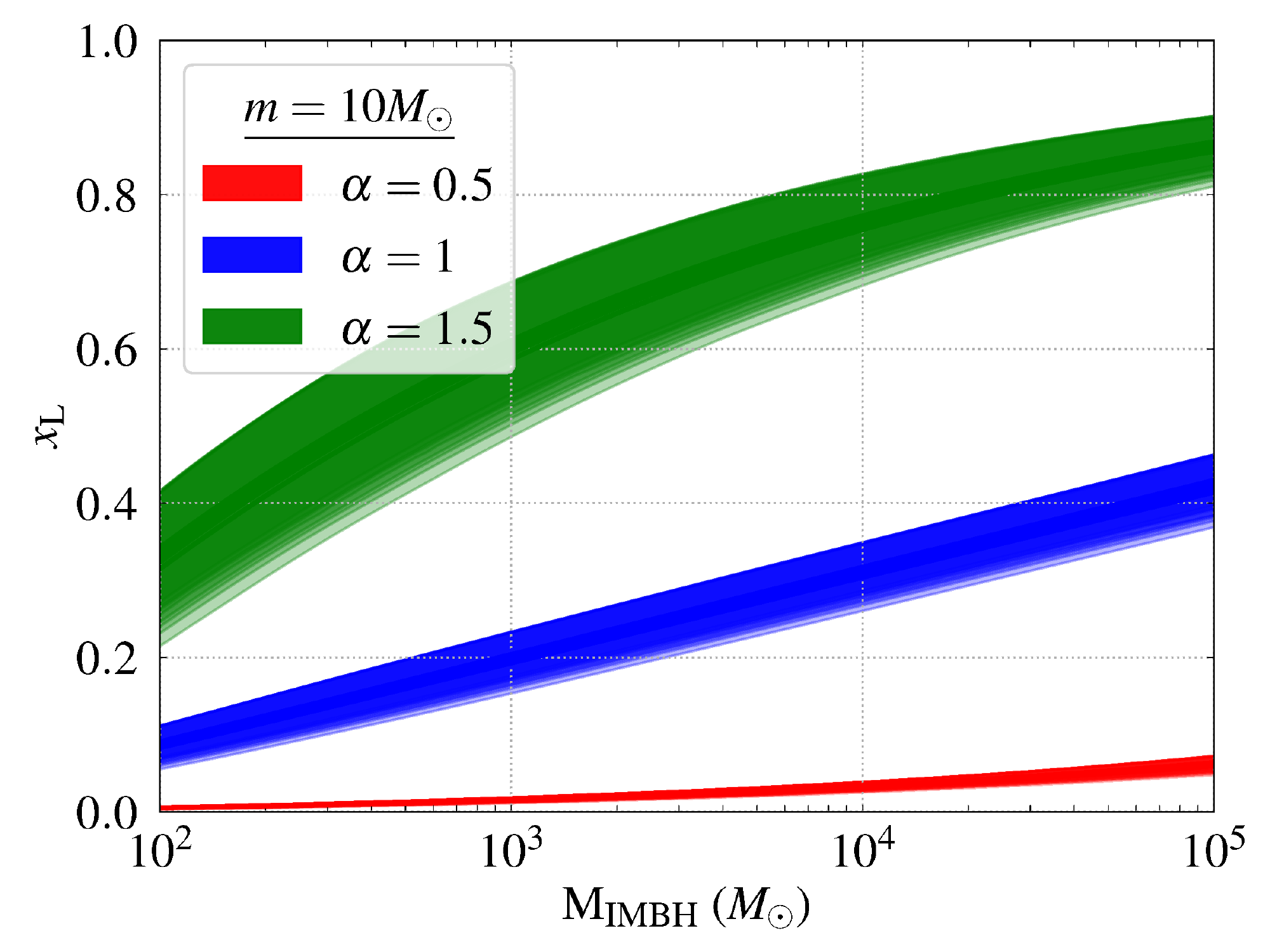}
  \caption{Fraction of IMRIs with detectable Doppler shifts for three values of the mass of the secondary companion: $m=0.6M_\odot$ (top), $m=1.4M_\odot$ (middle), and $m=10M_\odot$ (bottom). The curves of different colour corresponds to different slopes of the power-law distribution of semimajor axis of the IMRIs: $\alpha=0.5$ (red), $\alpha=1$ (blue), and $\alpha=1.5$ (green). The shaded regions show the scatter due to different distances to Sgr~A$^\star$ as well as random orientations of the orbits (both the IMRI's and the IMRI--Sgr~A$^\star$ orbit).\label{fig:fractions}}%
\end{figure}

Figure~\ref{fig:fractions} shows the fraction of detectable IMRIs as a function of the IMBH mass. The three panels correspond to the different choices for the mass of the secondary (top: $m=0.6\,M_\odot$; centre: $m=1.4\,M_\odot$; bottom: $m=10\,M_\odot$), while the results for different slopes of the power-law distribution of the semimajor axes are reported in different colours (red: $\alpha=0.5$; blue: $\alpha=1$; green: $\alpha=1.5$). The shaded regions show the scatter due to the combined effect of the different distances~$r$, different orientations of the IMRI's orbit, and random IMRI--Sgr~A$^\star$ inclinations. In more detail, for each~$r$ from the range under consideration, its shaded region shows the r.m.s. deviation from the mean resulting from all the random orientations of the orbits. As these individual regions are plotted on top of each other, the scatter increases even further to include the effect of different distances of the IMRI from the central SMBH.

It is evident from the graphs that the major factors contributing to the measurement of the Doppler-shifted GW signal from an IMRI are the IMRI's semimajor axis (i.e., the value of~$\alpha$) and the IMBH mass. The orbital orientations, the mass of the secondary companion, and the distance to Sgr~A$^\star$ have a rather mild effect on the detectable fraction. The weak dependence of the result on the mass of the secondary can be attributed to the small mass ratio. For $m\ll M_{\rm IMBH}$, the chirp mass is $\mathcal{M}\approx M_{\rm IMBH}(m/M_{\rm IMBH})^{3/5}$, thus SNR~$\propto\mathcal{M}^{5/6}=M_{\rm IMBH}^{1/3}m^{1/2}$. Everything else being equal, a change from a~WD to a~BH companion increases the SNR by about a factor of four, which may modify the cutoff value~$a_{\rm SNR}$. However, the resulting change in the fraction of detectable IMRIs turns out to be comparable to that due to random orbital orientations. There is a noticeable increase in the detectable fraction for the lightest IMBHs with BH companions (see Fig.~\ref{fig:fractions}, bottom panel), but it still remains below $\approx 40\%$. The IMBH mass in the SNR pre-factor does not have a significant effect either, but it affects the GW frequency (see Eq.~(\ref{eq:fIMRI})) and can significantly increase the SNR by bringing the binary to a more sensitive part of the LISA frequency band. The strong effect of the slope and the weak effect of the distance are consistent with our previous considerations. Indeed, the more IMRIs with $a\sim 0.01\;\mbox{AU}$, the more likely they are to be detected, since they emit in a GW frequency band where LISA is more sensitive (see Eqs.~(\ref{eq:aIMRI}) and~(\ref{eq:fIMRI})). In addition, if the IMRIs are located at $0.1$--$2\;\mbox{mpc}$ from the central SMBH, the RV variations occur fast enough to be detected within the LISA observation time (Eq.~(\ref{eq:TRV})). In other words, \textit{as long as an IMRI at a distance between~0.1~mpc and~2~mpc is detected, its RV is most likely measurable.}

\section{Scenarios to form intermediate-mass ratio inspirals}
\label{sec:scenarios}

In this section, we discuss the various channels to form an IMRI in the Galactic Centre, and compare different timescales relevant to the formation process.

\subsection{Model of the Galactic Centre}

We start off with a model of the Galactic nuclear star cluster (NSC) \citep[e.g.,][]{Hopman:2006xn, Alexander:2008tq,Gondan:2017wzd}. The model comprises four populations: MSs ($m_{\rm MS}=1\,M_\odot$), WDs ($m_{\rm WD}=0.6\,M_\odot$), NSs ($m_{\rm NS}=1.4\,M_\odot$), and BHs ($m_{\rm BH}=10\,M_\odot$). The approximate ratios of their total abundances within a radius of influence~$r_{\rm h}$ are correspondingly $1:0.1:0.01:0.001$, while their number densities follow cuspy power-law profiles~\citep{Bahcall:1976aa}
\be
n_{i} = n_{0,i}\left(\frac{r}{r_{\rm h}}\right)^{-\alpha_i}\,,
\ee
where $i$ enumerates the species ($i=\mbox{MS},\mbox{NS},\mbox{WD},\mbox{BH}$), $r_{\rm h}$ is the gravitational influence radius of Sgr~A$^\star$, $n_{0,i}$ are the respective number densities at that radius, and the slopes are $\alpha_{\rm MS}=\alpha_{\rm WD}=1.4$, $\alpha_{\rm NS}=1.5$, $\alpha_{\rm BH}=2$, with mass segregation being responsible for the steeper slopes~\citep{Hopman:2006xn}. The influence radius is defined as $r_{\rm h} = GM_{\rm SMBH}/v_{\rm h}^2$, with the velocity dispersion~$v_{\rm h}$ determined from the empirical $M$--$v_{\rm h}$ relation\,\footnote{We use~$v_{\rm h}$ for the velocity dispersion, and reserve the letter~$\sigma$ for cross sections below.}~\citep{Tremaine:2002js}:
\be
M_{\rm SMBH} \simeq 1.3\times 10^8M_\odot\,\left(\frac{v_{\rm h}}{200\;\mbox{km/s}}\right)^4\,,
\ee
which results in
\be
r_{\rm h} \simeq 2.4\;\mbox{pc}\,\sqrt{\frac{M_{\rm SMBH}}{4\times 10^6M_\odot}}\,.
\label{eq:infl_radius}
\ee
Note that there is no equipartition in the NSC, and the velocity dispersion is expected to be similar for all the sub-populations~\citep{Alexander:2000yb}. In what follows, we adopt $M_{\rm SMBH}\approx 4\times 10^6M_\odot$ for the mass of the SMBH in the centre of the Milky Way~\citep[e.g.,][]{Ghez:2008ms,Gillessen:2009ht,EventHorizonTelescope:2022wkp}.

To calibrate the central number density~$n_{\rm 0,MS}$ for MSs, we assume that the mass in stars within the influence radius is double the mass of Sgr~A$^\star$, \mbox{$M_\star(r<r_{\rm h})=2M_{\rm SMBH}$}~\citep{2013degn.book.....M}. This condition by definition implies a total of $N_{\rm MS}=8\times 10^6$ solar-mass stars, thus (neglecting the contribution of the compact remnants)
\be
\frac{n_{\rm 0,MS}}{3-\alpha_{\rm MS}} \approx 4.4\times 10^4\;\mbox{pc}^{-3}\,.
\ee
Regarding the number density of BHs, we normalise it following estimates that suggest $\approx 20,000$ BHs in the Galactic Centre~\citep[e.g.,][]{Freitag:2006qf,Hopman:2006xn}. This leads to a relative abundance of~$\sim 10^{-3}$, as in~\cite{Alexander:2008tq}, and 
\be
\frac{n_{\rm 0,BH}}{3-\alpha_{\rm BH}} \approx 2.5\times 10^{-3}\,\frac{n_{\rm 0,MS}}{3-\alpha_{\rm MS}}\,.
\ee
Finally, for WDs and NSs their central concentrations are set to
\be
n_{\rm 0,WD}=0.1n_{\rm 0,MS}\,, \qquad \frac{n_{\rm 0,NS}}{3-\alpha_{\rm NS}} \approx 10^{-2}\,\frac{n_{\rm 0,MS}}{3-\alpha_{\rm MS}}\,.
\ee

We also estimate the minimum radius~$r_{\rm min}$ (with respect to the SMBH) at which the continuous approximation for the distributions of MSs and compact objects breaks down. We define this radius as the distance at which the Poisson fluctuations in the numbers of objects~$N_i$ become comparable to the numbers themselves. Using a threshold of~$10\%$ for the relative fluctuation, we find for the radii encompassing $\sim 100$~MSs or $100$~BHs
\beqa
r_{\rm min,MS} &\simeq& 8.6\times 10^{-4} r_{\rm h} \approx 2\;\mbox{mpc}\,, \\
r_{\rm min,BH} &\simeq& 5\times 10^{-3} r_{\rm h} \approx 12\;\mbox{mpc}\,.
\eeqa
Note that this distance is of the order of the semimajor axes of the S-stars (and/or of their closest approach to Sgr~A$^\star$). For example, for S2, which is one of the closest stars to the central SMBH, the semimajor axis and minimum distance to Sgr~A$^\star$ are $\approx 5$~mpc and~$\approx 0.6$~mpc, respectively~\citep{Gillessen:2009ht}.

\subsection{Timescales}
\label{subsec:timescales}

If an IMBH is initially located on the outskirts of the sphere of influence, it starts sinking towards Sgr~A$^\star$ due to dynamical friction~\citep{1943ApJ....97..255C}. When it reaches the region $r\lesssim r_{\rm min}$, dynamical friction becomes inefficient and the slingshot effect takes over~\citep[e.g.,][]{Begelman:1980vb,Milosavljevic:2002bn}, with the SMBH--IMBH binary hardening as it interacts with and ejects stars coming from the bulk of the NSC~\citep[e.g.,][]{Quinlan:1996vp,Levin:2005dj,Sesana:2006xw,2019ApJ...878...17R}. Finally, the SMBH--IMBH separation becomes small enough for the GW emission to start dominating over the slingshot, and the binary undergoes a GW inspiral and merges.

While sinking towards the central SMBH, the IMBH may form a binary with a star or a stellar remnant. Possible channels for binary formation are gravitational captures of single BHs or NSs, exchanges with BH-BH binaries, and three-body interactions with single BHs and NSs. Here, we exclude from our analysis some interactions either because the respective timescales exceed the Hubble time (such as in the case of exchanges with NS-NS binaries) or because a consistent treatment would require taking into account tidal interactions (such as in the case of three-body interactions with WDs). We now compute and compare the timescales for dynamical friction and slingshot hardening to the timescales for the formation of an IMRI and the GW emission timescale.

\begin{itemize}

\item \textit{Dynamical friction}. The dependence of this timescale on the distance to Sgr~A$^\star$ is given by~\citep{AtakanGurkan:2004qm,Fragione:2020rmf}
\be
T_{\rm DF} = \left(1+\frac{4-\alpha}{6-\alpha}\zeta\right)\frac{(1+\zeta)^{1/2}}{(3-\alpha)BQ\zeta}\frac{r_{\rm h}}{v_{\rm h}}\left(\frac{r}{r_{\rm h}}\right)^{3/2}\,,
\ee
where $\alpha\approx 1.6$ is an effective slope of the cusp, $\zeta(r)=M_\star(r)/M_{\rm SMBH}$ is the mass of stars in units of the central SMBH's mass, $Q=M_{\rm IMBH}/M_{\rm SMBH}$ is the mass ratio, and $B\approx 1.7$ is a factor which absorbs the Coulomb logarithm and the ratio of the local circular velocity to velocity dispersion. The above formula is valid within a factor of a~few, and corrections may result from the contribution of the other objects (other than stars) as well as from varying the effective slope. Note that in numerical simulations it was also found that the sinking occurs on a somewhat longer timescale than suggested by analytical estimates~\citep[e.g.,][]{Baumgardt:2006jj}.

\item \textit{Slingshots and GW inspiral}. Since the transition from the slingshot hardening to the GW inspiral is expected to occur in the region of interest ($0.1\;\mbox{mpc}\lesssim r\lesssim 1\;\mbox{mpc}$), the timescales for the two processes should be treated jointly. If each timescale is estimated as the inverse of the logarithmic hardening rate,
\be
\label{eq:individual}
T_{\rm H} = \frac{r}{\dot{r}_{\rm H}}\,, \qquad T_{\rm GW} = \frac{r}{\dot{r}_{\rm \scriptscriptstyle GW}}\,,
\ee
then for the combined timescale we obtain~\citep[e.g.,][]{Sesana:2015haa}:
\beqa
\frac{1}{T_{\rm tot}} = \frac{\dot{r}}{r} &=& \frac{\dot{r}_{\rm H}}{r} + \frac{\dot{r}_{\rm \scriptscriptstyle GW}}{r} = \frac{1}{T_{\rm H}} + \frac{1}{T_{\rm GW}}\,, \nonumber \\
T_{\rm tot} &=& \frac{T_{\rm H}T_{\rm GW}}{T_{\rm H} + T_{\rm GW}}\,. \label{eq:combined}
\eeqa 

Now, regarding the individual timescales, the slingshot effect switches on at $r\lesssim 10a_{\rm H}$, where $a_H=GM_{\rm IMBH}/(4v^2)$~\citep{Sesana:2006xw}, with $v$ being the velocity dispersion far from the SMBH--IMBH binary (``at infinity''). In the range $0.01<a/a_{\rm H}<100$, the slingshot hardening timescale is~\citep{2019ApJ...878...17R}
\be
\label{eq:slingshot}
T_{\rm H} \approx 13\;\mbox{Myr}\,\frac{v}{100\;\mbox{km/s}}\left(\frac{H}{17}\right)^{-1} \times\left(\frac{\rho}{10^5\;M_\odot\,\mbox{pc}^{-3}}\right)^{-1}\left(\frac{r}{1\;\mbox{mpc}}\right)^{-1}\,,
\ee
where $\rho$ is the mass density of stars at infinity and $H$ is the hardening rate
\be
H(r) = 16.8\left(1 + \frac{r/a_{\rm H}}{3.21}\right)^{-0.73}\,.
\ee
Note that this fit is valid in the range~$0.01<r/a_{\rm H}<100$ and for a mass ratio $Q=10^{-3}$, but we also use it for other mass ratios, since the dependence of~$H$ on the mass ratio is weak.

The GW timescale in the leading PN order reads~\citep{Peters:1963ux,Peters:1964zz}
\beqa
T_{\rm GW} &=& \frac{5}{64}\frac{r/c}{Q(1+Q)}\left(\frac{r}{GM_{\rm SMBH}/c^2}\right)^3 \nonumber \\
&\approx& 3.6\;\mbox{Myr}\,\left(\frac{r}{0.1\;\mbox{mpc}}\right)^4\left(\frac{M_{\rm SMBH}}{4\times 10^{6}M_\odot}\right)^{-3}\left(\frac{Q}{10^{-3}}\right)^{-1} \nonumber \\
{} \label{eq:GWinspiral}
\eeqa
for~$Q<<1$ ($c$ is the speed of light). Recall also that we are considering only circular orbits here (eccentricity is discussed in Section~\ref{sec:discussion}).

Another way to look at the combined timescale of the slingshots and GW inspiral is to imagine an ensemble of NSCs identical to the one in our Galaxy and containing identical IMBHs at random distances~$r$. The steady-state distribution in $\log{r}$ is then proportional to $1/\left(\log{r}\right)\dot{}=r/\dot{r}=T_{\rm tot}$ (and in the distances themselves, to $1/\dot{r}=T_{\rm tot}/r$). The maximum of this distribution (the dashed--dotted curves in Fig.~\ref{fig:timescales}) does not dramatically differ from a distance where $\dot{r}_{\rm H}=\dot{r}_{\rm\scriptscriptstyle GW}$~\citep[see also][]{Sesana:2015haa} or from the maximum of function~$T_{\rm tot}(r)/r$.%

\item \textit{Gravitational captures}. An IMBH can capture an initially unbound compact remnant if enough energy is radiated away in GWs. The GW capture cross section~$\sigma_{\rm cap}$ is given in terms of a maximum impact parameter~$b_{\rm cap}$ as follows~\citep{1987ApJ...321..199Q,OLeary:2008myb}:
\be
\sigma_{\rm cap} = \pi b_{\rm cap}^2\,, \quad b_{\rm cap} = \frac{\sqrt{2G R_{\rm cap}M_{\rm IMBH}(1+q)}}{v}\,,
\ee
where $q\equiv m/M_{\rm IMBH}$ is the mass ratio of the IMRI, and the distance of closest approach
\beqa
R_{\rm cap} &=& \frac{2GM_{\rm IMBH}}{c^2}\left(\frac{85\pi}{96}\right)^{2/7}\left(\frac{c}{v}\right)^{4/7}q^{2/7}(1+q)^{3/7} \nonumber \\
&\simeq& 5000\;\mbox{km}\,\left(\frac{M_{\rm IMBH}}{10^{3}M_\odot}\right)\left(\frac{v_{\rm h}}{100\;\mbox{km/s}}\right)^{-4/7}\nonumber \\
&\times&\left(\frac{r}{1\;\mbox{mpc}}\right)^{2/7}\left(\frac{q}{10^{-3}}\right)^{2/7}\,,
\eeqa
where we have approximated the velocity dispersion $v\sim\sqrt{GM_{\rm SMBH}/r}$. From this characteristic value of the closest approach, it is clear that this channel can only work for either BHs or NSs. For both MSs and WDs, tidal effects will come into play long before GW emission becomes important. The timescale for gravitational capture is then given by \citep{Fragione:2020rmf}
\be
T_{\rm cap} = \frac{1}{n\sigma_{\rm cap}v}\,,
\ee
where $n$ stands for the number density of either BHs or NSs.

\begin{figure*}
  \centering
  \includegraphics[width=0.49\textwidth]{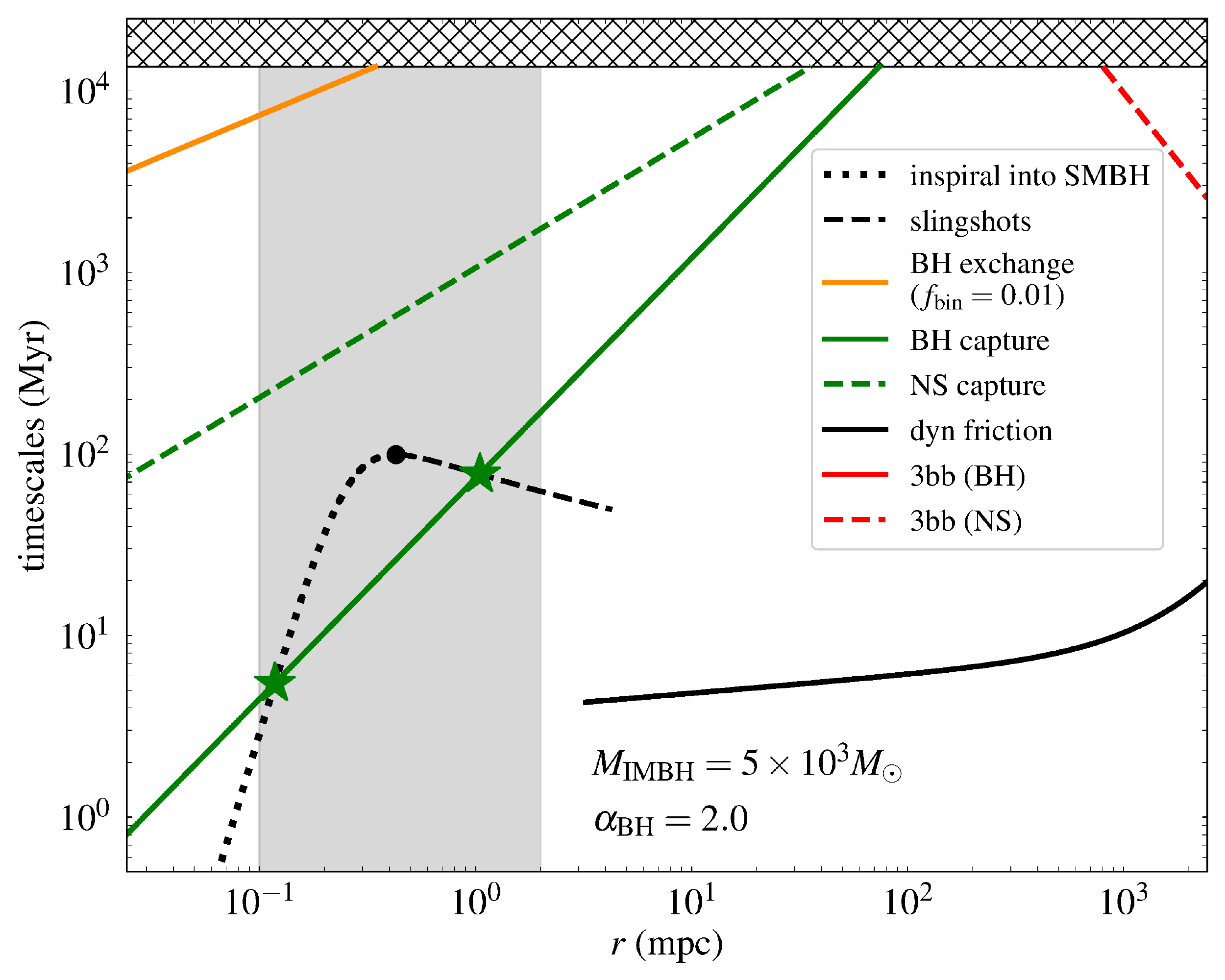}
  \includegraphics[width=0.49\textwidth]{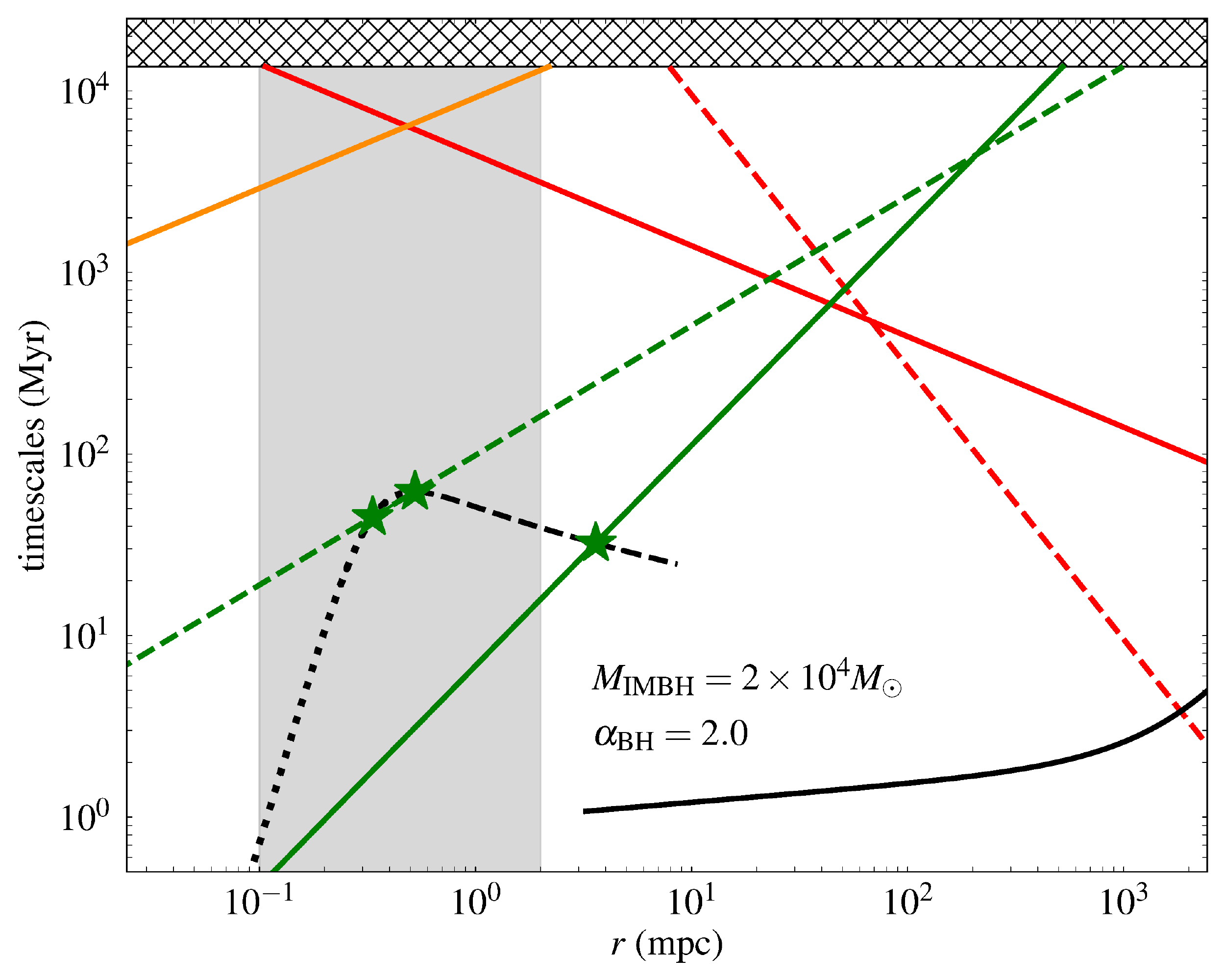}
  \includegraphics[width=0.49\textwidth]{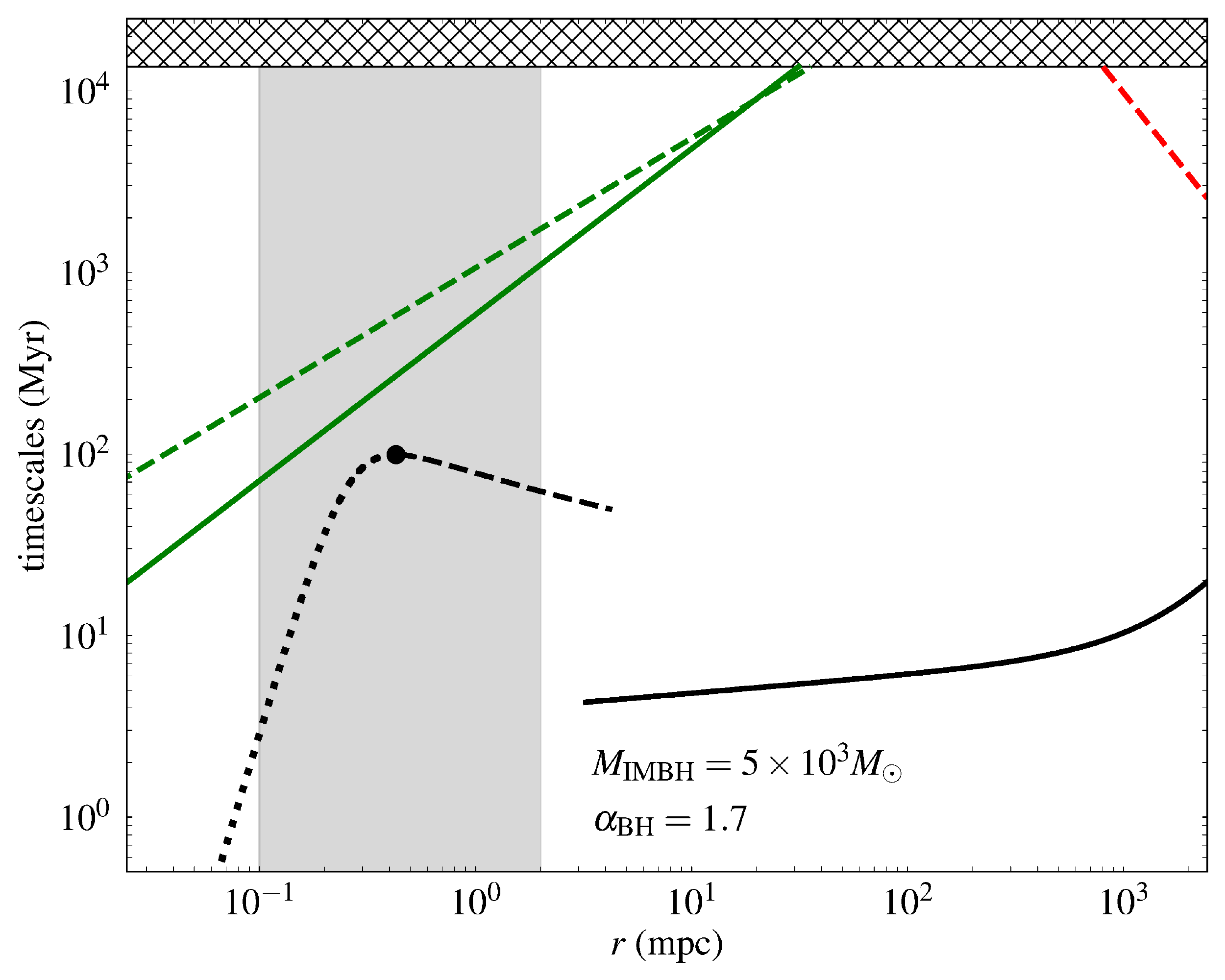}
  \includegraphics[width=0.49\textwidth]{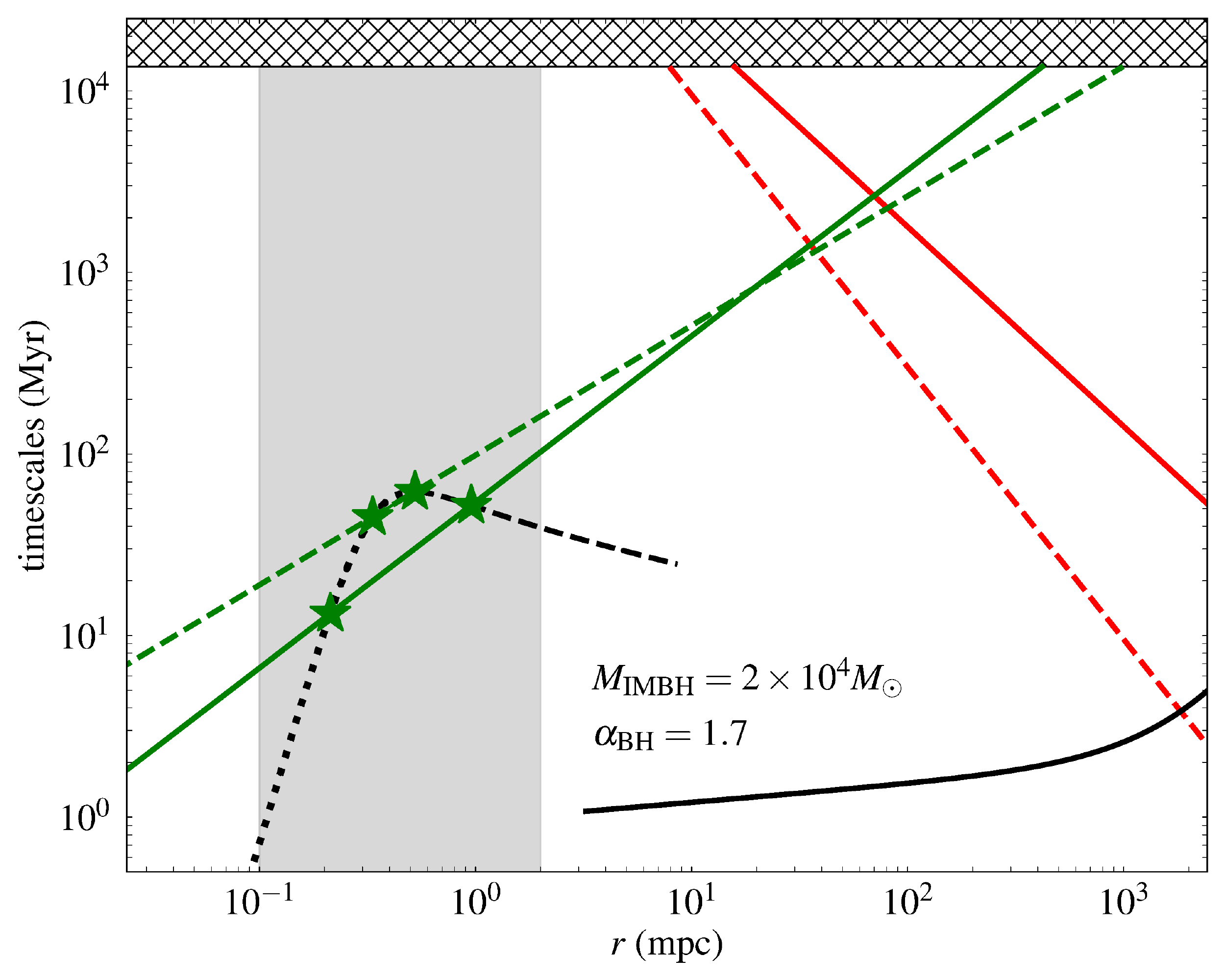}
  \caption{Timescales for the inspiral of an IMBH toward the central SMBH and for formation of an IMRI. Columns correspond to two cases of the IMBH mass (left: $M_{\rm IMBH}=5\times 10^{3}M_\odot$; right: $M_{\rm IMBH}=2\times 10^{4}M_\odot$), while rows show the timescales for two values of the power-law slope of the BH cusp (top: $\alpha_{\rm BH}=2$; bottom: $\alpha_{\rm BH}=1.7$). The black lines show different stages of the IMBH inspiral: dynamical friction (solid line), slingshots (dashed), and GW~emission (dotted). Transition from the slingshots to the GW inspiral occurs in the region of interest ($0.1\;\mbox{mpc}<r<2\;\mbox{mpc}$; grey shaded area), and the respective timescales are combined to highlight the transition, Eqs.~(\ref{eq:combined})--(\ref{eq:GWinspiral}). The maximum of the combined timescale is marked with a black dot. The solid line is cut off at a radius where the dynamical friction stalls and the slingshot effect takes over. The green lines show the timescale for gravitational capture of BHs (solid line) and NSs (dashed). As soon as this timescale becomes shorter than the combined slingshot--GW emission timescale (green stars), the gravitational captures can proceed efficiently. Other timescales are for BH binary disruption (orange) and three-body interaction with BHs (red, solid) and NSs (red, dashed). The range of~$r$ on all panels correspond to $10^{-5}<r/r_{\rm h}<1$, where $r_{\rm h}$ is the influence radius, Eq.~(\ref{eq:infl_radius}). The hatched area corresponds to times longer than the Hubble time.\label{fig:timescales}}%
\end{figure*}

\item \textit{Exchanges}. When an IMBH interacts with a binary BH or NS, it can capture one of the binary's components, with the other one escaping to infinity. To survive in the NSC, the binary must be ``hard'' with a separation $a_{\rm bin}\lesssim Gm/(4v^2)$, where $m=m_{\rm BH}=10M_\odot$ or $m=m_{\rm NS}=1.4M_\odot$. Such a binary is prone to disruption by the IMBH if it comes closer than the tidal disruption limit~\citep{2001MNRAS.321..398M}
\be
b_{\rm ex} \simeq 4a_{\rm bin}\left(\frac{M_{\rm IMBH}}{m_{\rm bin}}\right)^{1/3}\,.
\ee
The characteristic timescale is then
\be
T_{\rm ex} = \frac{1}{n_{\rm bin}(\pi b_{\rm ex}^2)v}\,,
\ee
where we assume that the number density of the binaries is a fraction $f_{\rm bin}$ of the number density of BHs or NSs, $n_{\rm bin}=f_{\rm bin}n$. In our calculation, we set $f_{\rm bin}=0.01$ for both, but this parameter is rather uncertain.

\item \textit{Three-body interactions}. It is also possible to form an IMRI through three-body interactions. The timescale for this process is essentially inversely proportional to the probability of finding two stellar-mass objects within the sphere of influence of the IMBH:
\be
T_{\rm 3bb} = K\frac{a_{\rm IMRI}/v}{(na_{\rm IMRI}^3)^2}\,,
\label{eq:3bb_timescale}
\ee
where $K$ is a proportionality factor. It can be estimated by looking at a similar expression for the case of three-body interactions of stellar BHs. For example, comparing Eq.~(\ref{eq:3bb_timescale}) to~Eq.~(2) of~\citet{Morscher:2014doa}, we see the same overall dependence on the mass and velocity, and we can estimate
\be
K \simeq  \frac{10^{-3}}{2^{10}(1+q)^5}
\ee
for an IMRI with a marginally hard separation.  

\end{itemize}

Figure~\ref{fig:timescales} shows the timescales for~$M_{\rm IMBH}=5\times 10^3M_\odot$ (left column) and~$M_{\rm IMBH}=2\times 10^4M_\odot$ (right column) and for two assumptions about the slope of the BH cusp (top: $\alpha_{\rm BH}=2$; bottom: $\alpha_{\rm BH}=1.7$). Firstly, as expected~\citep[e.g.,][]{Hansen:2003yb,Arca-Sedda:2018kne,Khan:2021jqf}, we obtain that an IMBH deposited at the outskirts of the NSC sinks to the region of interest in~$\lesssim 100$~Myr. If we combine this with the rate at which the IMBH deposition occurs, we can estimate how likely it is to find an IMBH in the Galactic Centre today. Both theoretical arguments and observations suggest that at least $50\%$ of the mass of the Milky Way's NSC was formed through infall of globular clusters~\citep[e.g.,][]{Antonini:2011xh, Neumayer:2020gno, Fahrion:2021wmz}. Considering that an IMBH typically forms within \mbox{$\sim 1$~Gyr}~\citep{2015MNRAS.454.3150G,2021ApJ...908L..29G}, which is also a typical time it takes for a cluster to sink from a radius~$\sim 1$~kpc to the Galactic Centre~\citep{Gnedin:2013cda}, we can expect $\sim 0.1$--$1\;\mbox{Gyr}^{-1}$ for the IMBH deposition rate in the Milky Way~\citep[see also][]{Arca-Sedda:2018kne}. By Little's law\,\footnote{Little's law/theorem in queueing theory states that the number of customers in a queue is the product of the time a customer spends in the queue with the arrival rate~\citep[see, for example,][]{1990psqt.book.....A}.}, this results in a probability $\lesssim (0.1\mbox{--}1\;\mbox{Gyr}^{-1})\times 100\;\mbox{Myr}=1$--$10\%$\,.

Secondly, from an overall comparison of the timescales in Fig.~\ref{fig:timescales}, GW captures are likely the most promising channel for the formation of an IMRI at the distances under consideration (the grey shaded bands in all panels). Captures can proceed efficiently as soon as their timescale becomes shorter than that due to slingshots and GW emission; the radii at which this happens are marked by green stars in Fig.~\ref{fig:timescales}. Stellar BHs are captured more easily when their cusp is steeper ($\alpha_{\rm BH}=2$, top), whereas the rates for BHs and NSs are comparable in a shallower cusp ($\alpha_{\rm BH}=1.7$, bottom). In a steeper cusp, GW captures of BHs occur at $r\simeq 0.1$--$1$~mpc, and the trend is that more massive IMBHs capture at larger radii. The timescale estimates suggest that captures by IMBHs with mass $M_{\rm IMBH}\gtrsim 10^4M_\odot$ happen at the transition from the dynamical friction to slingshot inspiral stage, while those by lighter IMBHs happen at the slingshot stage. The GW captures of NSs are somewhat efficient at smaller radii in a steeper cusp, $r\approx 0.1$--$1$~mpc, and for heavier IMBHs with mass~$M_{\rm IMBH}\gtrsim 10^4M_\odot$. More massive IMBHs can capture a NS at the transition from the slignshot hardening to the GW inspiral.
In a shallower cusp and for heavier IMBHs, the BH captures are brought down to approximately the same radii as the NS captures (as we changed the slope of the BH cusp but kept the slope for NSs fixed). Lighter IMBHs with mass~$M_{\rm IMBH}\lesssim 10^4M_\odot$ do not appear to efficiently capture either BHs or NSs in shallower cusps.

Regarding the other processes, while the binary formation through exchanges is negligible in a shallower cusp, this process may play a role in a steeper cusp. Even in the case of a more massive IMBH, $T_{\rm ex}\sim 1$--$10$~Gyr, which is too long for the exchanges to significantly contribute to IMRI formation. However, the trend is that this timescale decreases as the IMBH mass increases. Another factor that can make it shorter is a higher fraction of BH-BH binaries. It is thus possible that the exchanges might start contributing for $M_{\rm IMBH}\gtrsim 10^5M_\odot$. The same applies to three-body interactions with stellar-mass BHs. As for three-body interactions with NSs, seemingly they may play a role for more massive IMBHs at larger radii (right column; the red dashed line intersecting with the black solid line). However, we caution that our estimates of the timescales were based on the assumption that the velocity dispersion is dominated by the central SMBH, which is no longer the case at distances comparable to its influence radius.

\section{Discussion}
\label{sec:discussion}

In this paper, we have argued that LISA could constrain the presence of an IMBH in the Galactic Centre if the IMBH forms a binary (an IMRI) with a compact remnant (a WD, a NS, or a BH), while orbiting Sgr~A$^\star$ at distances $\sim 0.1$--$1$~mpc. First, we have done order-of-magnitude estimates to show that, if such a binary is detected, the Doppler shift in its GW signal is almost necessarily detected as well. We have then confirmed those estimates with a Fisher matrix analysis of the RV uncertainties, and we have found that IMRIs with an IMBH of $10^3$--$10^5M_\odot$ and separations $\sim 0.01$--$0.1$~AU are most likely to exhibit Doppler-shift variations detectable by LISA. 

For a power-law distribution of the separations $\propto a^{-3/2}$, this results in a detectability fraction of $50\%$--$80\%$ for WD and NS companions and $60\%$--$85\%$ for BHs, with larger fractions corresponding to heavier IMBHs. This fraction is only weakly affected by the distance to the central SMBH as long as $0.1\;\mbox{mpc}<r<2\;\mbox{mpc}$. The overall uncertainty on the fraction is $\pm 10\%$ ($\pm 5\%$ for the heaviest IMBHs with $M_{\rm IMBH}\sim 10^5M_\odot$) and accounts for both the varying distance and random inclinations of the orbits. The respective percentages for the log-uniform distribution ($\propto a^{-1}$) and for an even shallower power law ($\propto a^{-1/2}$) are $15\%$--$40\%$ (with errors~$\pm 5\%$) and $1\%$--$5\%$ (with errors~$\pm 1\%$). Therefore, if an IMRI with mass $10^3M_\odot$--$10^5M_\odot$ is in the LISA band, its radial velocity is measurable. What fraction of the IMRIs end up in the LISA band is defined by the distribution of IMRI separations.

We have also discussed the various channels that could lead to the formation of an IMRI. We have found that GW captures of BHs and NSs are not only the most efficient way to form an IMRI at the distances of interest, but they also almost inevitably produce an IMRI \citep[see also][]{Fragione:2020rmf}. The captures continue to be more or less efficient in both steeper and shallower BH cusps and for heavier or lighter IMBHs. All in all, this and the considerations on the Doppler shift detectability imply that LISA can provide additional constraints on a potential IMBH in our Galactic Centre.

One limitation of this study is the assumption of circular orbits for our Fisher matrix analysis. For example, the ``outer'' IMRI--Sgr~A$^\star$ binary may acquire an eccentricity~$e_{\rm out}$ as it sinks due to dynamical friction~\citep{Baumgardt:2006jj}. With the Milky Way NSC rotating as a whole~\citep[][Section~5.6 and references therein]{Neumayer:2020gno}, $e_{\rm out}$~may also exhibit a qualitatively different behaviour for prograde and retrograde ``outer'' orbits~\citep{Khan:2021jqf}. In particular, the prograde orbits tend to go through a shorter slingshot phase and be more circular when entering the GW emission phase, while the retrograde ones have longer slinghot timescales but acquire significant eccentricities and experience a prompt GW merger~\citep[see e.g. the simulation of M32 by][]{Khan:2021jqf}.

Moderate eccentricities~$e_{\rm out}$ are not expected to alter the RV detectability. To reiterate, RV variations are detectable as long as they occur on a timescale shorter than the observation time, and, if the semimajor axis is maintained fixed, the eccentricity does not change the timescale (given by the orbital period). However, the eccentricity will affect the inspiral time of the IMBH into the central SMBH as well as the stability of the IMRI if it comes too close to the SMBH at pericentre. For instance, for $e_{\rm out}=0.9$, the dashed-boundary grey region in Fig.~\ref{fig:constraints} would correspond to an inspiral time $<3$~Myr rather than $10$~Myr, and an IMRI with mass~$\gtrsim 10^3M_\odot$ and a semimajor axis $\sim 0.01$--$0.1$~AU would not undergo von Zeipel--Kozai--Lidov (ZKL) oscillations at the closest approach \citep{2016ARA&A..54..441N}. Neither effect is expected to be significant whenever $e_{\rm out}\lesssim 0.9$. Note also that, as the orbit becomes more eccentric, relativistic corrections to the Doppler shift may start playing a role and can be used to break degeneracies between orbital parameters of the triple system~\citep{2022arXiv221209753K}. For $e_{\rm out}=0.9$ and a semimajor axis of~$0.1$~mpc, the distance of closest approach is of the order of about $30$~Schwarzschild radii of the central SMBH (assuming it is non-spinning). These effects come into play for $e_{\rm out}\gtrsim 0.99$; however, in this case the GW emission of the IMBH--Sgr~A$^\star$ binary would also peak at a frequency that falls well within the LISA band\,\footnote{To calculate the peak GW frequency~$f_{\rm peak}$, we use a fit by~\citet{Wen:2002km}: $f_{\rm peak}=2f_0(1+e)^{1.1954}/(1-e)$, where $f_0$ is the orbital frequency and $e$, the eccentricity. For $e=0.99$ and a semimajor axis~$0.1$~mpc, this gives $f_{\rm peak}\simeq 1$~mHz.}.

An eccentricity~$e_{\rm in}$ of the ``inner binary'' IMRI may have opposite effects on the detectability. On one hand, the eccentricity could enhance the GW signal and shift it to a more sensitive region of the LISA band, thus making the source ``brighter''~\citep{1963PhRv..131..435P}. On the other hand, eccentric IMRIs inspiral faster and make it less likely for LISA to spot one during the observation time. Similarly to the case discussed in the previous paragraph, for moderate eccentricities ($e_{\rm in}\lesssim 0.9$) the effect of the shorter inspiral time is not important: it is reduced only by $1/3$ for $e_{\rm in}=0.9$. In fact, for this value the peak GW frequency of an IMRI with mass $10^3M_\odot$ and semimajor axis~$0.01$~AU would shift to where LISA is somewhat more sensitive, from $2$~mHz to $50$~mHz. Such shifts are not quite favourable for heavier IMBHs. High intrinsic eccentricities result in short-lived IMRIs. As we discussed in Sec.~\ref{subsec:timescales}, IMRIs in the Galactic Centre may effectively be produced via GW captures, which would naturally result in very eccentric binaries. For example, for $M_{\rm IMBH}=5\times 10^3M_\odot$ and~$M_{\rm IMBH}=2\times 10^4M_\odot$, the distributions of semimajor axis, $(1-e_{\rm in}^2)$, GW frequency, and inspiral time are peaked correspondingly at: $2$~AU and $20$~AU, $\sim 10^{-4}$ for both, $0.2$~Hz and~$0.05$~Hz, and $0.3$~yr and~$10$~yr \citep{Fragione:2020rmf}. Although these IMRIs are still in the LISA band, the short inspiral times make their detection unlikely. On a technical note, apart from the assumption about circular orbits, other assumptions underlying our calculation break down for these highly eccentric IMRIs as well. Namely, it was assumed that the increase in GW frequency happens on a timescale much longer than both the IMRI's orbital period and the orbital period around the central SMBH. All three timescales become comparable in the case of IMRIs formed via GW captures. A comprehensive treatment of this case may call for a full parameter estimation, which we leave to future work.

Since the LISA constraints discussed in this paper apply mostly to quite massive IMBHs of $\sim 10^3M_\odot$--$10^5M_\odot$ at relatively small distances $\sim 0.1$--$1$~mpc (see Fig.~\ref{fig:constraints}), the question arises whether those IMBHs leave any trail as they sink to the Galactic Centre. Simulations of IMBH--IMBH mergers in dwarf galaxies show that, during the slingshot phase, the total mass of ejected stars is $\approx\mbox{a few}\times M_{\rm IMBH}$~\citep{Khan:2021jqf}. On one hand, the erosion caused by a sinking IMBH may be masked by recent star formation that occurs on a timescale of \mbox{$\lesssim 100$~Myr}~\citep[e.g.][]{Walcher:2006hd}. This scenario is quite plausible, given that the in-situ star formation may be responsible for up to~$50\%$ of the NSC's stellar mass~\citep{Fahrion:2021wmz}. On the other hand, the ejected stars are expected to have high velocities~\citep{2019ApJ...878...17R} and may be spotted as the so-called hypervelocity stars~\citep{1988Natur.331..687H,2005ApJ...622L..33B}, as recently discussed by~\cite{2023arXiv230412169E}.

To conclude, our proposed way of putting additional constraints on an IMBH in the Galactic Centre is probabilistic: the robustness of the constraints depends on how likely it is to have an IMRI at the distances of interest. Our consideration of such a possibility was based on the comparison of timescales for different formation channels. A more precise study of the binary formation should include $N$-body simulations of IMBHs in NSCs, including the possibility that they could be delivered by infalling star clusters \citep[e.g.,][]{Arca-Sedda:2018kne}. However, a trend we saw is that heavier IMBHs in a steeper BH cusp are more likely to end up in a binary. Therefore, LISA is more useful to constrain IMBHs of $\sim 10^4M_\odot$--$10^5M_\odot$ at $\sim 0.1$--$1$~mpc, complementing present and upcoming constraints.

\section*{Acknowledgements}
V.S. and E.B. were supported by NSF Grants No. PHY-1912550 and AST-2006538, NASA ATP Grants No. 17-ATP17-0225 and 19-ATP19-0051, NSF-XSEDE Grant No. PHY-090003, and NSF Grant PHY-20043. G.F., V.S., and E.B. acknowledge support from NASA Grant 80NSSC21K1722.
G.F. acknowledges support from NSF Grant~AST-1716762 at Northwestern University. The authors are grateful to the anonymous referee for a constructive report and for drawing our attention to details of the hardening process for a SMBH--IMBH binary. We would also like to thank Bence Kocsis for pointing out to us additional relevant references.
The authors acknowledge the Texas Advanced Computing Center (TACC) at The University of Texas at Austin for providing HPC resources that have contributed to the research results reported within this paper~\citep{10.1145/3311790.3396656}
(URL: \url{http://www.tacc.utexas.edu}).

\textit{Software}. IPython~\citep{2007CSE.....9c..21P}, SciPy~\citep{2020NatMe..17..261V},  Matplotlib~\citep{2007CSE.....9...90H}, NumPy~\citep{2011CSE....13b..22V}, SymPy~\citep{Meurer:2017yhf}, \texttt{mpmath}~\citep{mpmath}, \texttt{filltex}~\citep{2017JOSS....2..222G}.

\section*{Data Availability}

The output of the simulations used to compute the detectability fraction is available upon a reasonable request to the corresponding author.

\bibliographystyle{mnras}
\bibliography{refs} %

\bsp	%
\label{lastpage}
\end{document}